\documentclass[12pt]{article}

\usepackage{a4,amsmath,amssymb}
\usepackage[hypertex]{hyperref}

\newcommand{\di}{\genfrac{}{}{0pt}{}}

\usepackage[arrow,curve,frame]{xy}
\makeatletter
\xydef@\double@#1{\edef\Drop@@{\dimen@=#1\relax
 \dimen@=1.5\dimen@ \A@=-\sinDirection\dimen@ \B@=\cosDirection\dimen@
 \setboxz@h{\setbox2=\hbox{\kern\A@\raise\B@\copy\z@}%
 \dp2=\z@ \ht2=\z@ \wd2=\z@ \box2
 \setbox2=\hbox{\kern-\A@\raise-\B@ \noexpand\boxz@}%
 \dp2=\z@ \ht2=\z@ \wd2=\z@ \box2 }%
 \ht\z@=\z@ \dp\z@=\z@ \wd\z@=\z@ \noexpand\styledboxz@}}

\def\section{\@startsection{section}{1}{\z@}{-3.25ex plus -1ex minus
    -.2ex}{1.5ex plus .2ex}{\normalfont\large\bfseries}}
\def\subsection{\@startsection{subsection}{1}{\z@}{-3.25ex plus -1ex
    minus -.2ex}{1.5ex plus .2ex}{\normalfont\itshape}}

\renewenvironment{thebibliography}[1]
         {\section*{References}\frenchspacing\small
          \begin{list}{[\arabic{enumi}]}
         {\usecounter{enumi}\parsep=2pt\topsep 0pt
         \settowidth{\labelwidth}{[#1]}
         \leftmargin=\labelwidth\advance\leftmargin\labelsep
         \rightmargin=0pt\itemsep=0pt\sloppy}}{\end{list}}

\makeatother

\renewcommand{\title}[1]{\vspace{10mm}\noindent{\Large{\bf #1}}\vspace{8mm}}
\newcommand{\authors}[1]{\noindent{\large #1}\vspace{3mm}}
\newcommand{\address}[1]{{\itshape #1\vspace{2mm}}}

\newtheorem{Theorem}{Theorem}
\begin{document}

\begin{titlepage}

\begin{center}

\title{Progress in solving a noncommutative \\[2mm]  quantum field theory 
in four dimensions}

\authors{Harald {\sc Grosse}$^1$ and Raimar {\sc Wulkenhaar}$^2$}

\address{$^{1}$\,Fakult\"at f\"ur Physik, Universit\"at Wien\\
Boltzmanngasse 5, A-1090 Wien, Austria}

\address{$^{2}$\,Mathematisches Institut der Westf\"alischen
  Wilhelms-Universit\"at\\
Einsteinstra\ss{}e 62, D-48149 M\"unster, Germany}

\footnotetext[1]{harald.grosse@univie.ac.at}
\footnotetext[2]{raimar@math.uni-muenster.de}

\vskip 3cm

\textbf{Abstract} \vskip 3mm
\begin{minipage}{14cm}%
  We study the noncommutative $\phi^4_4$-quantum field theory at the
  self-duality point. This model is renormalisable to all orders as
  shown in earlier work of us and does not have a Landau ghost
  problem. Using the Ward identity of Disertori, Gurau, Magnen and
  Rivasseau, we obtain from the Schwinger-Dyson equation a non-linear
  integral equation for the renormalised two-point function alone. The
  non-trivial renormalised four-point function fulfils a linear
  integral equation with the inhomogeneity determined by the two-point
  function. These integral equations are the starting point for a
  perturbative solution. In this way, the renormalised correlation
  functions are directly obtained, without Feynman graph computation and 
  further renormalisation steps. 
\end{minipage}

\end{center}
\end{titlepage}

\section{Introduction}

In order to improve the problems of four-dimensional quantum field
theory it was suggested to include ``gravity effects'' through
deforming space-time. The canonical deformation is particularly
simple, but the resulting models suffer from the UV/IR-mixing
\cite{Minwalla:1999px}.

In our previous work \cite{Grosse:2004yu} we found a way to handle
this problem. We realised that the model defined by the action
\begin{align}
S=\int d^4x\Big(
\frac{1}{2} \phi (-\Delta+ \Omega^2 \tilde{x}^2 +\mu^2) \phi
+ \frac{\lambda}{4} \phi\star \phi\star \phi \star \phi \Big)(x)
\label{GW}
\end{align}
is renormalisable to all orders of perturbation theory. Here, $\star$
refers to the Moyal product parametrised by the antisymmetric $4\times
4$-matrix $\Theta$, and $\tilde{x}=2\Theta^{-1}x$.  The model is
covariant under the Langmann-Szabo duality transformation
\cite{Langmann:2002cc} and becomes self-dual at $\Omega=1$. Certain
variants have also been treated, see \cite{Rivasseau:2007ab} for a
review.

Evaluation of the $\beta$-functions for the coupling constants
$\Omega,\lambda$ in first order of perturbation theory leads to a
coupled dynamical system which indicates a fixed-point at $\Omega=1$,
while $\lambda$ remains bounded \cite{Grosse:2004by,Grosse:2005da}.
The vanishing of the $\beta$-function at $\Omega=1$ was next proven in
\cite{Disertori:2006uy} at three-loop order and finally in
\cite{Disertori:2006nq} to all orders of perturbation theory. It
implies that there is no infinite renormalisation of $\lambda$, and a
non-perturbative construction seems possible
\cite{Rivasseau:2007fr}. The Landau ghost problem is solved.

The vanishing of the $\beta$-function to all orders has been obtained
using a Ward identity \cite{Disertori:2006nq}. We extend this work and
derive an integral equation for the two-point function alone by using
the Ward identity and Schwinger-Dyson equations. Usually,
Schwinger-Dyson equations couple the two-point function to the
four-point function. In our model, we show that the Ward identity
allows to express the four-point function in terms of the two-point
function, resulting in an equation for the two-point function
alone. This is achieved in the first step for the bare two-point
function. We are able to perform the mass and wavefunction
renormalisation directly in the integral equation, giving a
\emph{self-consistent non-linear equation for the renormalised
  two-point function alone}.

Higher $n$-point functions fulfil a \emph{linear} (inhomogeneous) 
Schwinger-Dyson equation, with the inhomogeneity given by $m$-point
functions with $m<n$. This means that solving our equation for the
two-point function leads to a full non-perturbative construction of
this interacting quantum field theory in four dimensions. 

So far we treated our equation perturbatively up to third order in
$\lambda$. The solution shows an interesting number-theoretic
structure. It takes values in a polynomial ring with generators
\begin{align}
\alpha,\ \beta,\ \frac{1-\alpha}{1-\alpha\beta},\  
\frac{1-\beta}{1-\alpha\beta},\ \{I_{t(\alpha)}\},\ \{I_{t(\beta)}\}
\end{align}
and rational coefficients, where the $I_{t(\alpha)}$ are iterated
integrals labelled by rooted trees. Similar structures also appeared
in toy models for the Connes-Kreimer Hopf algebra
\cite{Connes:1998qv}.  The $I_{t(\alpha)}$ evaluate to polylogarithms
and zeta functions \cite{Kontsevich:2001}.

We hope that a detailed analysis of our model will help for a
non-perturbative treatment of more realistic quantum field theories.

\section{Action functional and Ward identity}

It is convenient to write the action (\ref{GW}) in the matrix base of
the Moyal space, see \cite{Grosse:2004yu,Grosse:2003nw}. It simplifies
enormously at the self-duality point $\Omega=1$. We write down the
resulting action functionals for the \emph{bare} quantities, which
involves the bare mass $\mu_{bare}$ and the wave function
renormalisation $\phi \mapsto Z^{\frac{1}{2}}\phi$. For simplicity
we fix the length scale to $\theta=4$. This gives
\begin{align}
S&=\sum_{m,n \in \mathbb{N}^2_\Lambda} 
\frac{1}{2} \phi_{mn} H_{mn} \phi_{nm} + V(\phi)\:,
\\
H_{mn}
&= Z\big(\mu_{bare}^2 + |m|+|n| \big) \;, 
\qquad
V(\phi)= \frac{Z^2\lambda}{4}\sum_{m,n,k,l \in \mathbb{N}^2_\Lambda} 
\phi_{mn} \phi_{nk} \phi_{kl} \phi_{lm}\;,
\end{align}
It is already used that this model has no renormalisation of the
coupling constant \cite{Disertori:2006nq}. All summation indices
$m,n,\dots$ belong to $\mathbb{N}^2$, with $|m|:=m_1+m_2$. The symbol
$\mathbb{N}^2_\Lambda$ refers to a cut-off in the matrix size. The
scalar field is real, $\phi_{mn}=\overline{\phi_{nm}}$.

We recall the derivation of the Ward identity from
\cite{Disertori:2006nq}.  We study a unitary transformation $\phi_{mn}
\mapsto \sum_{k,l \in \mathbb{N}^2_\Lambda} U_{mk} \phi_{kl} U^\dag_{ln}$ and
its infinitesimal version
\begin{align}
\phi_{mn}
\mapsto \phi_{mn}+ \mathrm{i} \sum_{k \in \mathbb{N}^2_\Lambda} 
(B_{mk} \phi_{kn} - \phi_{mk} B_{kn})\;.
\end{align}
In contrast to the action functional, the partition function
\begin{align}
\mathcal{Z}[J]=N \int \mathcal{D} \phi \; e^{-S+\mathrm{tr} (\phi J)}
\end{align}
will be invariant under such a transformation. 
The measure is $\mathcal{D} \phi=\prod_{m,n \in \mathbb{N}^2_\Lambda}
d \phi_{mn}$, again with cut-off in the matrix size.
The trace is given by 
$\mathrm{tr} (\phi J)=\sum_{k,l \in \mathbb{N}^2_\Lambda} \phi_{kl} J_{lk}$.
We consider the variation of the generating functional $W=\ln \mathcal{Z}$ of
connected functions:
\begin{align}
0 &= \frac{\delta W}{\mathrm{i}\delta B_{ab}} 
= \frac{1}{\mathcal{Z}}\int \mathcal{D} \phi \; 
\Big(-\frac{\delta S}{\mathrm{i}\delta B_{ab}} 
+ \frac{\delta}{\mathrm{i}\delta B_{ab}}\big(  \mathrm{tr} (\phi J)\big)\Big)
e^{-S+ \mathrm{tr} (\phi J)}
\nonumber
\\
&= \frac{1}{\mathcal{Z}}\int \mathcal{D} \phi \; 
\sum_{n} \Big( 
(H_{nb}-H_{an}) \phi_{bn} \phi_{na}
+ (\phi_{bn} J_{na} - J_{bn} \phi_{na}) 
\Big) e^{-S+\mathrm{tr} (\phi J)}\;.
\end{align}
In the perturbative expansion, the fields in interaction vertices are 
written as derivatives with respect to the sources, $\phi_{mn} \mapsto
\frac{\delta}{\delta J_{nm}}$. After functional integration, we obtain
the Ward identity 
\begin{align}
0 &= \Big\{ \sum_{n} \Big(
(H_{nb}-H_{an}) \frac{\delta^2 }{\delta J_{nb}\,\delta J_{an}}
+  \Big(J_{na} \frac{\delta}{\delta J_{nb}} - J_{bn}
\frac{\delta}{\delta J_{an}}\Big) 
\Big) \nonumber \\
& \qquad \times \exp\Big(-V\big(\tfrac{\delta}{\delta J}\big)\Big)
e^{\frac{1}{2} \sum_{p,q} J_{pq} H^{-1}_{pq} J_{qp}}\Big\}_{c}\;.
\label{Ward-J}
\end{align}
Only the connected functions (symbolised by the subscript $c$) are
generated. The Ward identity (\ref{Ward-J}) tells us that inserting 
into the connected graphs one special insertion vertex
\begin{align}
\label{V-ins}
V^{ins}_{ab} :=  \sum_n (H_{an}-H_{nb}) \phi_{bn}\phi_{na}
\end{align}
is the same as the difference between the exchanges of external 
sources $J_{nb} \mapsto J_{na}$ and $J_{an} \mapsto J_{bn}$.

We write Feynman graphs in the Langmann-Szabo self-dual
$\phi^4_4$-model as ribbon graphs on a genus-$g$ Riemann surface with
$B$ external faces.  Adding for each external face an external vertex
to get a closed surface, the matrix index is constant at every
face. Inserting the special vertex $V^{ins}_{ab}$ leads, however, to
an index jump from $a$ to $b$ in an external face which meets an external
vertex. The corresponding external sources at the jumped face are thus
$J_{na}$ and $J_{bm}$ for some other indices $m,n$.  According to the
Ward identity, this is the same as the difference between the graphs
with face index $b$ and $a$, respectively:
\begin{align}
Z (|a|-|b|)\parbox{35mm}{\begin{picture}(35,20)
\put(2,9){{\xy <1cm,0cm>:
(0,0) *++={ }; (1,0) *++={ } **\frm{o},
(0.41,0.41)="Bi",
(1.11,1.11)="Bo",
(0.41,-0.41)="Ci",
(1.21,-1.11)="Co",
(-.39,0.41)="Ai",
(-1.1,1.11)="Ao",
(-0.39,-0.41)="Di",
(-1.1,-1.2)="Do",
(1.1,0.65)="D1",
(1.1,-0.65)="D2",
(-1.5,0.04)="E1",
(-1.5,-0.04)="E2",
(-1.58,0.12)="E1a",
(-1.58,-0.12)="E2a",
(-1.65,0)="E0",
\ar @{=} "Bi";"Bo"
\ar @{{ }<} "Bi";"Bo" <3pt>
\ar @{{ }>} "Bi";"Bo" <-3pt>
\ar @{=} "Ci";"Co"
\ar @{{ }<} "Ci";"Co" <3pt>
\ar @{{ }>} "Ci";"Co" <-3pt>
\ar @`{(1.8,0)} @{{ }*=<2mm>{ .}} "D1";"D2"
\ar @`{"Do"} @{=} "Di";"E2"
\ar @`{"Ao"} @{=} "Ai";"E1"
\ar @{<{ }} "E1";"Ao" <3pt>
\ar @{<{ }} "E2";"Do" <3pt>
\ar @`{"E0"} @{.} "E1a";"E2a"
\endxy}}
\put(1,12){\mbox{\scriptsize$a$}}
\put(1,4){\mbox{\scriptsize$b$}}
\put(25,20){\mbox{\scriptsize$a$}}
\put(25,-2){\mbox{\scriptsize$b$}}
\end{picture}} &=
\parbox{26mm}{\begin{picture}(26,20)
\put(7,9){{\xy <1cm,0cm>:
(0,0) *++={ }; (1,0) *++={ } **\frm{o},
(0.41,0.41)="Bi",
(1.11,1.11)="Bo",
(0.41,-0.41)="Ci",
(1.21,-1.11)="Co",
(1.1,0.65)="D1",
(1.1,-0.65)="D2",
\ar @{=} "Bi";"Bo"
\ar @{{ }<} "Bi";"Bo" <3pt>
\ar @{{ }>} "Bi";"Bo" <-3pt>
\ar @{=} "Ci";"Co"
\ar @{{ }<} "Ci";"Co" <3pt>
\ar @{{ }>} "Ci";"Co" <-3pt>
\ar @`{(1.8,0)} @{{ }*=<2mm>{ .}} "D1";"D2"
\endxy}}
\put(14,19){\mbox{\scriptsize$b$}}
\put(14,-1){\mbox{\scriptsize$b$}}
\end{picture}}
-
\parbox{26mm}{\begin{picture}(26,20)
\put(7,9){{\xy <1cm,0cm>:
(0,0) *++={ }; (1,0) *++={ } **\frm{o},
(0.41,0.41)="Bi",
(1.11,1.11)="Bo",
(0.41,-0.41)="Ci",
(1.21,-1.11)="Co",
(1.1,0.65)="D1",
(1.1,-0.65)="D2",
\ar @{=} "Bi";"Bo"
\ar @{{ }<} "Bi";"Bo" <3pt>
\ar @{{ }>} "Bi";"Bo" <-3pt>
\ar @{=} "Ci";"Co"
\ar @{{ }<} "Ci";"Co" <3pt>
\ar @{{ }>} "Ci";"Co" <-3pt>
\ar @`{(1.8,0)} @{{ }*=<2mm>{ .}} "D1";"D2"
\endxy}}
\put(14,19){\mbox{\scriptsize$a$}}
\put(14,-1){\mbox{\scriptsize$a$}}
\end{picture}}
\\[3ex]
Z(|a|-|b|) G^{ins}_{[ab]\dots} &= 
G_{b\dots}-G_{a\dots}\;.
\label{WI-ins}
\end{align}
The dots in (\ref{WI-ins}) stand for the remaining face indices.
We have used $H_{an}-H_{nb}=Z(|a|-|b|)$.

\section{Two-point Schwinger-Dyson equation}

We consider the Schwinger-Dyson equation for the one-particle 
irreducible (1PI) \emph{planar} two-point function with 
respect to the leftmost vertex:
\begin{align}
\underbrace{\parbox{30mm}{\begin{picture}(30,20)
\put(0,10){\xy <1cm,0cm>:
(0,0) *++={ }; (1,0) *++={ } **\frm{oo},
(0.55,0)="Bi",
(1.3,0)="Bo",
(-0.55,0)="Ai",
(-1.3,0)="Ao",
\ar @{:} "Ai";"Ao"
\ar @{{ }<} "Ai";"Ao" <3pt> 
\ar @{{ }>} "Ai";"Ao" <-3pt>
\ar @{:} "Bi";"Bo"
\ar @{{ }<} "Bi";"Bo" <3pt> 
\ar @{{ }>} "Bi";"Bo" <-3pt>
\endxy}
\put(2,13){\mbox{\scriptsize$a$}}
\put(2,6){\mbox{\scriptsize$b$}}
\put(22,13){\mbox{\scriptsize$a$}}
\put(22,6){\mbox{\scriptsize$b$}}
\end{picture}}}_{\Gamma_{ab}}
&= 
\underbrace{\parbox{25mm}{\begin{picture}(25,30)
\put(3,14){\xy <1cm,0cm>:
(0.4,0.8) *++={ }; (0.9,0.8) *++={ } **\frm{o},
(-0.8,0)="Ai",
(-1.5,0)="Ao",
(-0.7,-0.1)="Di",
(-0.7,-0.8)="Do",
(-0.7,0.1)="Fi",
(-0.7,0.8)="Fm",
(0.05,0.8)="Fo",
(-0.6,0)="Gi",
(0.4,0)="Gm",
(0.4,0.45)="Go",
\ar @{:} "Ai";"Ao"
\ar @{{ }<} "Ai";"Ao" <3pt> 
\ar @{{ }>} "Ai";"Ao" <-3pt>
\ar @{:} "Di";"Do"
\ar @{{ }<} "Di";"Do" <3pt> 
\ar @{{ }>} "Di";"Do" <-3pt>
\ar @`{"Fm"} @{=} "Fi";"Fo"
\ar @`{"Gm"} @{=} "Gi";"Go"
\endxy}
\put(8,7){\mbox{\scriptsize$b$}}
\put(14,7){\mbox{\scriptsize$a$}}
\put(4,16){\mbox{\scriptsize$a$}}
\put(4,10){\mbox{\scriptsize$b$}}
\end{picture}}}_{T^L_{ab}}
+  \underbrace{\parbox{25mm}{\begin{picture}(25,30)
\put(3,14){\xy <1cm,0cm>:
(0.4,-0.8) *++={ }; (0.9,-0.8) *++={ } **\frm{o},
(-0.8,0)="Ai",
(-1.5,0)="Ao",
(-0.7,0.1)="Di",
(-0.7,0.8)="Do",
(-0.7,-0.1)="Fi",
(-0.7,-0.8)="Fm",
(0.05,-0.8)="Fo",
(-0.6,0)="Gi",
(0.4,0)="Gm",
(0.4,-0.45)="Go",
\ar @{:} "Ai";"Ao"
\ar @{{ }<} "Ai";"Ao" <3pt> 
\ar @{{ }>} "Ai";"Ao" <-3pt>
\ar @{:} "Di";"Do"
\ar @{{ }<} "Di";"Do" <3pt> 
\ar @{{ }>} "Di";"Do" <-3pt>
\ar @`{"Fm"} @{=} "Fi";"Fo"
\ar @`{"Gm"} @{=} "Gi";"Go"
\endxy}
\put(14,18){\mbox{\scriptsize$b$}}
\put(8,18){\mbox{\scriptsize$a$}}
\put(4,16){\mbox{\scriptsize$a$}}
\put(4,10){\mbox{\scriptsize$b$}}
\put(14,10){\mbox{\scriptsize$p$}}
\end{picture}}
+ 
\parbox{40mm}{\begin{picture}(20,20)
\put(3,10){\xy <1cm,0cm>:
(0.75,0) *++={ }; (1.5,0) *++={ } **\frm{oo},
(1.15,0)="Bi",
(1.9,0)="Bo",
(-0.8,0)="Ai",
(-1.5,0)="Ao",
(-0.7,-0.1)="Di",
(-0.7,-0.6)="Do",
(-0.7,0.1)="Fi",
(-0.7,0.7)="Fm",
(0.2,0.7)="Fn",
(0.5,0.35)="Fo",
(-0.7,-0.1)="Di",
(-0.7,-0.9)="Dm",
(0.2,-0.8)="Dn",
(0.52,-0.33)="Do",
(-0.6,0)="Gi",
(0.35,0)="Gm",
\ar @{:} "Ai";"Ao"
\ar @{{ }<} "Ai";"Ao" <3pt> 
\ar @{{ }>} "Ai";"Ao" <-3pt>
\ar @{:} "Bi";"Bo"
\ar @{{ }<} "Bi";"Bo" <3pt> 
\ar @{{ }>} "Bi";"Bo" <-3pt>
\ar @`{"Dm","Dn"} @{=} "Di";"Do"
\ar @`{"Fm","Fn"} @{=} "Fi";"Fo"
\ar @{=} "Gi";"Gm"
\endxy}
\put(5,12){\mbox{\scriptsize$a$}}
\put(5,6){\mbox{\scriptsize$b$}}
\put(36,13){\mbox{\scriptsize$a$}}
\put(36,6){\mbox{\scriptsize$b$}}
\put(14,6){\mbox{\scriptsize$p$}}
\end{picture}}}_{\Sigma^R_{ab}}
\label{TSigma}
\end{align}
A double circle in (\ref{TSigma}) stands for 1PI subgraphs, a single
circle for connected graphs.  In the graphs contributing to 
$\Sigma^R_{ab}$ we open the $p$-face
and compare it with the insertion into the connected two-point
function. There are two different places of an insertion: either into
a one-particle-\emph{reducible} propagator, or into an 1PI two-point
function:
\begin{align}
G^{ins}_{[ap]b} &= 
\parbox{32mm}{\begin{picture}(32,20)
\put(2,9){{\xy <1cm,0cm>:
(0,0) *++={ }; (1,0) *++={ } **\frm{o},
(0.46,0.31)="Bi",
(1.1,0.81)="Bo",
(0.46,-0.31)="Ci",
(1.1,-0.81)="Co",
(-.39,0.41)="Ai",
(-1.1,1.11)="Ao",
(-0.39,-0.41)="Di",
(-1.1,-1.2)="Do",
(-1.5,0.04)="E1",
(-1.5,-0.04)="E2",
(-1.58,0.12)="E1a",
(-1.58,-0.12)="E2a",
(-1.65,0)="E0",
\ar @{=} "Bi";"Bo"
\ar @{{ }<} "Bi";"Bo" <3pt>
\ar @{{ }>} "Bi";"Bo" <-3pt>
\ar @{=} "Ci";"Co"
\ar @{{ }<} "Ci";"Co" <3pt>
\ar @{{ }>} "Ci";"Co" <-3pt>
\ar @`{"Do"} @{=} "Di";"E2"
\ar @`{"Ao"} @{=} "Ai";"E1"
\ar @{<{ }} "E1";"Ao" <3pt>
\ar @{<{ }} "E2";"Do" <3pt>
\ar @`{"E0"} @{.} "E1a";"E2a"
\endxy}}
\put(1,12){\mbox{\scriptsize$a$}}
\put(1,4){\mbox{\scriptsize$p$}}
\put(26,18){\mbox{\scriptsize$a$}}
\put(28,12){\mbox{\scriptsize$b$}}
\put(25,-2){\mbox{\scriptsize$p$}}
\put(28,4){\mbox{\scriptsize$b$}}
\end{picture}} 
=
\parbox{30mm}{\begin{picture}(30,20)
\put(2,9){{\xy <1cm,0cm>:
(-0.05,0.8) *++={ }; (.55,0.8) *++={ } **\frm{o},
(-0.05,-0.8) *++={ }; (.55,-0.8) *++={ } **\frm{o},
(0.3,0.8)="Fi",
(0.9,0.8)="Fo",
(0.3,-0.8)="Gi",
(0.95,-0.8)="Go",
(-.38,0.91)="Ai",
(-1.1,1.11)="Ao",
(-0.38,-0.91)="Di",
(-1.1,-1.2)="Do",
(-1.5,0.04)="E1",
(-1.5,-0.04)="E2",
(-1.58,0.12)="E1a",
(-1.58,-0.12)="E2a",
(-1.65,0)="E0",
\ar @{=} "Fi";"Fo"
\ar @{{ }<} "Fi";"Fo" <3pt>
\ar @{{ }>} "Fi";"Fo" <-3pt>
\ar @{=} "Gi";"Go"
\ar @{{ }<} "Gi";"Go" <3pt>
\ar @{{ }>} "Gi";"Go" <-3pt>
\ar @`{"Do"} @{=} "Di";"E2"
\ar @`{"Ao"} @{=} "Ai";"E1"
\ar @{<{ }} "E1";"Ao" <3pt>
\ar @{<{ }} "E2";"Do" <3pt>
\ar @`{"E0"} @{.} "E1a";"E2a"
\endxy}}
\put(1,12){\mbox{\scriptsize$a$}}
\put(1,4){\mbox{\scriptsize$p$}}
\put(25,19){\mbox{\scriptsize$a$}}
\put(25,12){\mbox{\scriptsize$b$}}
\put(25,-3){\mbox{\scriptsize$p$}}
\put(25,4){\mbox{\scriptsize$b$}}
\end{picture}} 
+
\parbox{45mm}{\begin{picture}(35,20)
\put(2,9){{\xy <1cm,0cm>:
(0,0) *++={ }; (1,0) *++={ } **\frm{oo},
(1.1,0.8) *++={ }; (1.7,0.8) *++={ } **\frm{o},
(1.1,-0.8) *++={ }; (1.7,-0.8) *++={ } **\frm{o},
(0.46,0.31)="Bi",
(0.83,0.59)="Bo",
(0.46,-0.31)="Ci",
(0.83,-0.59)="Co",
(1.5,0.8)="Fi",
(2.05,0.8)="Fo",
(1.5,-0.8)="Gi",
(2.05,-0.8)="Go",
(-.39,0.41)="Ai",
(-1.1,1.11)="Ao",
(-0.39,-0.41)="Di",
(-1.1,-1.2)="Do",
(-1.5,0.04)="E1",
(-1.5,-0.04)="E2",
(-1.58,0.12)="E1a",
(-1.58,-0.12)="E2a",
(-1.65,0)="E0",
\ar @{=} "Bi";"Bo"
\ar @{=} "Ci";"Co"
\ar @{=} "Fi";"Fo"
\ar @{{ }<} "Fi";"Fo" <3pt>
\ar @{{ }>} "Fi";"Fo" <-3pt>
\ar @{=} "Gi";"Go"
\ar @{{ }<} "Gi";"Go" <3pt>
\ar @{{ }>} "Gi";"Go" <-3pt>
\ar @`{"Do"} @{=} "Di";"E2"
\ar @`{"Ao"} @{=} "Ai";"E1"
\ar @{<{ }} "E1";"Ao" <3pt>
\ar @{<{ }} "E2";"Do" <3pt>
\ar @`{"E0"} @{.} "E1a";"E2a"
\endxy}}
\put(1,12){\mbox{\scriptsize$a$}}
\put(1,4){\mbox{\scriptsize$p$}}
\put(38,19){\mbox{\scriptsize$a$}}
\put(38,12){\mbox{\scriptsize$b$}}
\put(38,-3){\mbox{\scriptsize$p$}}
\put(38,4){\mbox{\scriptsize$b$}}
\end{picture}} 
\end{align}
We amputate the upper $G_{ab}$ two-point function and sum over $p$.
After multiplication by the vertex $Z^2 \lambda$, the result 
is precisely the combination $\Sigma^R_{ab}$ of graphs:
\begin{align}
\Sigma^R_{ab} 
&= Z^2 \lambda \sum_p (G_{ab})^{-1} G^{ins}_{[ap]b} 
= - Z\lambda \sum_p (G_{ab})^{-1} \frac{G_{bp}-G_{ba}}{|p|-|a|} \;.
\label{SigmaRab}
\end{align}
The last step follows from (\ref{WI-ins}).  The special case $a=b=0$
and $Z=1$ of (\ref{SigmaRab}) already appeared in
  \cite{Disertori:2006nq}. \emph{The fact that we obtained this
    formula for all $a,b \in \mathbb{N}^2$ allows us to derive a
    Schwinger-Dyson equation (\ref{SD-2}) which involves only the
    two-point function, not the four-point function as usual.}  Noting
  that
\begin{align}
G_{ab}^{-1}=H_{ab}-\Gamma_{ab}
\end{align}
and $T^L_{ab}=Z^2\lambda \sum_q G_{aq}$ in (\ref{TSigma}), we have 
for the connected function 
\begin{align}
Z^2 \lambda \sum_q G_{aq} 
- Z\lambda \sum_p (G_{ab})^{-1} \frac{ G_{bp}-G_{ba}}{|p|-|a|}  
 = H_{ab}-G_{ab}^{-1}\;.
\label{SD-2}
\end{align}
We stress that the two-point function is by definition symmetric,
$\Gamma_{ab}=\Gamma_{ba}$, although this is not manifest in (\ref{SD-2})!

We express this Schwinger-Dyson equation in terms of the 1PI function
$\Gamma_{ab}$, because renormalisation is performed in the 1PI
part. After rearranging of $1=G_{ab}^{-1}G_{ba}
= G_{bp} G_{pb}^{-1}$, we have 
\begin{align}
\Gamma_{ab} &=  
Z^2 \lambda \sum_p \Big(
\frac{1}{H_{bp} -\Gamma_{bp}}
+\frac{1}{H_{ap} -\Gamma_{ap}}
-
\frac{1}{H_{bp} -\Gamma_{bp}} 
\frac{(\Gamma_{bp}-\Gamma_{ab})}{Z(|p|-|a|)}\Big)\;.
\end{align}
To pass to renormalised quantities, we Taylor expand
\begin{align}
\Gamma_{ab}= Z \mu_{bare}^2 -\mu^2 
+ (Z-1) (|a|+|b|) +\Gamma^{ren}_{ab}\;, 
\\
\Gamma^{ren}_{00}=0 \qquad (\partial \Gamma^{ren})_{00}=0\:,
\end{align}
where $\partial \Gamma^{ren}$ is any of the derivatives with respect to 
$a_1,a_2,b_1,b_2$. This implies
\begin{align}
G_{ab}^{-1}= |a|+|b|+\mu^2-\Gamma_{ab}^{ren}\;.
\label{G-inverse}  
\end{align}
Hence, $\mu$ is the renormalised mass, and both $G_{ab}$ and
$\Gamma_{ab}$ should be regular if the cut-off in the matrix indices
is removed.  The resulting equation is
\begin{align}
  &Z \mu_{bare}^2 -\mu^2 +
  (Z-1) (|a|+|b|) +\Gamma^{ren}_{ab}
  \nonumber
  \\
  &= \lambda  \sum_p \Big( 
\frac{Z}{|b|+|p|+\mu^2 -\Gamma^{ren}_{bp}} 
+\frac{Z^2}{|a|+|p|+\mu^2-\Gamma^{ren}_{ap}}
  \nonumber
  \\
  &- \frac{Z}{|b|+|p|+\mu^2 -\Gamma^{ren}_{bp}}
  \frac{\Gamma^{ren}_{bp}-\Gamma^{ren}_{ab}}{(|p|-|a|)}\Big)\;.
\label{Gamma-total}
\end{align}
Notice the difference of the exponent of $Z$ in the two tadpoles! 
Separating the first Taylor term we obtain
\begin{align}
Z \mu_{bare}^2 -\mu^2 
  &= \lambda \sum_p \Big( 
\frac{Z^2+Z}{|p|+\mu^2 -\Gamma^{ren}_{0p}} 
- \frac{Z}{|p|+\mu^2 -\Gamma^{ren}_{0p}}
  \frac{\Gamma^{ren}_{0p}}{|p|}\Big)
\end{align}
and 
\begin{align}
  &
  (Z-1) (|a|+|b|) +\Gamma^{ren}_{ab}
  \nonumber
  \\
  &= \lambda \sum_p \Big( 
\frac{Z}{|b|+|p|+\mu^2 -\Gamma^{ren}_{bp}} 
+\frac{Z^2}{|a|+|p|+\mu^2-\Gamma^{ren}_{ap}}
-\frac{Z^2+Z}{|p|+\mu^2-\Gamma^{ren}_{0p}}
  \nonumber
  \\
  &
- \frac{Z}{|b|+|p|+\mu^2 -\Gamma^{ren}_{bp}}
  \frac{\Gamma^{ren}_{bp}-\Gamma^{ren}_{ab}}{|p|-|a|}
+ \frac{Z}{p+\mu^2 -\Gamma^{ren}_{0p}}
  \frac{\Gamma^{ren}_{0p}}{|p|}\Big)\;.
\label{ZG}
\end{align}
Deriving (\ref{ZG}) at $0$ with respect to $a_i$ and $b_i$ leads to a
self-consistent system of equations for $Z,\Gamma_{ab}^{ren}$. In the
next section we analyse this system for continuous indices $a,b \in 
\mathbb{R}_+ \times \mathbb{R}_+$.

\section{Integral representation}

For simplicity we replace the indices in $\mathbb{N}$ by continuous
variables in $\mathbb{R}_+$.  It is crucial that (\ref{ZG}) depends
only on the sums $|a|=a_1+a_2$, $|b|=b_1+b_2$ and $|p|=p_1+p_2$ of
indices. Therefore, also the two-point function $\Gamma^{ren}_{ab}$
must depend on these sums only. This means that the sum $\sum_{p_1,p_2
  \in \mathbb{N}_\Lambda}$ is replaced by the integral
$\int_{0}^\Lambda |p| d|p|$, where we already introduced a cut-off
$|p|=p_1+p_2 \leq \Lambda$.  Instead of (\ref{ZG}) we thus have
\begin{align}
&(Z-1)(|a|+|b|) +\Gamma^{ren}_{ab}
\nonumber
\\
&= \int_0^\Lambda |p|\,d|p| \Big( 
\frac{Z}{|b|+|p|+\mu^2 -\Gamma^{ren}_{bp}} 
+\frac{Z^2}{|a|+|p|+\mu^2-\Gamma^{ren}_{ap}}
-\frac{Z^2+Z}{|p|+\mu^2-\Gamma^{ren}_{0p}}
  \nonumber
  \\
  &
\qquad - \frac{Z}{|b|+|p|+\mu^2 -\Gamma^{ren}_{bp}}
  \frac{\Gamma^{ren}_{bp}-\Gamma^{ren}_{ab}}{(|p|-|a|)}
+ \frac{Z}{|p|+\mu^2 -\Gamma^{ren}_{0p}}
  \frac{\Gamma^{ren}_{0p}}{|p|}
\Big)\;,
\end{align}
with $|a|,|b|,|p| \in \mathbb{R}_+$.
We introduce a change of variables 
\begin{align}
|a| &=:\mu^2\frac{\alpha}{1-\alpha}\;,\quad
|b|=:\mu^2 \frac{\beta}{1-\beta}\;,\quad
|p|=:\mu^2 \frac{\rho}{1-\rho}\;,\quad
|p|\,d|p| = \mu^4 \frac{ \rho\, d\rho}{(1-\rho)^3}
\nonumber
\\
\Gamma^{ren}_{ab} &=:\mu^2 
\frac{\Gamma_{\alpha\beta}}{(1-\alpha)(1-\beta)}\;,
\quad
\Lambda=:\mu^2 \frac{\xi}{1-\xi}
\label{alpha}
\end{align}
and obtain
\begin{align}
&(Z-1)\Big(\frac{\alpha}{1-\alpha}+\frac{\beta}{1-\beta}\Big) 
+\frac{\Gamma_{\alpha\beta}}{(1-\alpha)(1-\beta)}
\nonumber
\\
&= \lambda \int_0^\xi \frac{\rho \,d\rho}{(1-\rho)^2} \Big( 
\frac{Z^2(1-\alpha)}{1-\alpha \rho -\Gamma_{\alpha\rho}} 
-\frac{Z^2}{1 -\Gamma_{0\rho}} \Big)
  \nonumber
  \\
  &
-\lambda \int_0^\xi \frac{d\rho}{(1-\rho)} 
\Big(\frac{Z (1-\Gamma_{\beta\alpha})
}{1-\beta \rho -\Gamma_{\beta\rho}} 
+ \frac{Z\alpha }{1-\beta \rho -\Gamma_{\beta\rho}} 
  \frac{\Gamma_{\beta\rho}-\Gamma_{\beta\alpha} }{\rho-\alpha}
- \frac{Z}{1-\Gamma_{0\rho}}
\Big)\;.
\label{Gamma-alphabeta}
\end{align}
We have $\frac{\partial}{\partial
  a_i}\big|_{a=0}=\frac{\partial}{\partial |a|}\big|_{a=0}
=(1-\alpha)^2\frac{\partial}{\partial\alpha}\big|_{\alpha=0}
=\frac{\partial}{\partial\alpha}\big|_{\alpha=0}$ 
so that we obtain with $ \Gamma'_{0\rho}:= \lim_{\alpha \to 0} 
\frac{\Gamma_{\alpha\rho}- \Gamma_{0\rho}}{\alpha}$
the following two relations for $Z$:
\begin{align}
Z-1&= -Z\lambda \int_0^\xi \frac{d\rho}{(1-\rho)} \;
\frac{(\rho + \Gamma'_{0\rho})}{(1 -\Gamma_{0\rho})^2} \;,
\label{Z1}
\\
Z-1&= Z^2\lambda \int_0^\xi \!\! \frac{\rho\,d\rho}{(1-\rho)^2} 
\Big( \frac{\rho+\Gamma'_{0\rho}}{(1-\Gamma_{0\rho})^2}
-\frac{1}{1-\Gamma_{0\rho}}\Big)
- Z\lambda \int_0^\xi \!\! \frac{d\rho}{(1-\rho)} 
\frac{1}{1-\Gamma_{0\rho}} \frac{\Gamma_{0\rho}}{\rho}\;.
\label{Z2}
\end{align}

We now express (\ref{Gamma-alphabeta}) in terms of the connected
function $G_{\alpha\beta}$ defined by
\begin{align}
1-\alpha\beta-\Gamma_{\alpha\beta}=\frac{1-\alpha\beta}{
G_{\alpha\beta}} \;.
\end{align}
The result is
\begin{align}
&Z G_{\alpha\beta} -1 
- (Z-1) \frac{(1-\alpha)(1-\beta)}{1-\alpha\beta} G_{\alpha\beta}
\nonumber
\\
&= \lambda Z^2 G_{\alpha\beta} \frac{(1-\alpha)(1-\beta)}{1-\alpha\beta}
\int_0^\xi \frac{\rho \,d\rho}{(1-\rho)^2} \Big( 
\frac{(1-\alpha)G_{\alpha\rho}}{1-\alpha \rho} 
- G_{0\rho} \Big)
  \nonumber
  \\
  &
+\lambda Z G_{\alpha\beta} 
\frac{(1-\alpha)(1-\beta)}{1-\alpha\beta}
\int_0^\xi \frac{d\rho}{(1-\rho)}  G_{0\rho}
  \nonumber
  \\
  &
-\lambda Z (1-\alpha)(1-\beta)
\int_0^\xi \frac{d\rho}{(1-\rho)} 
\Big(\frac{\rho}{1-\beta \rho} \frac{G_{\beta\rho}}{\rho-\alpha}
- \frac{\alpha}{1-\beta \alpha} \frac{G_{\beta\alpha}}{\rho-\alpha}
\Big)\;.
\label{G-alphabeta}
\end{align}
Using $\rho+\Gamma_{0\rho}'=
\frac{\rho}{G_{0\rho}}+\frac{G_{0\rho}'}{G_{0\rho}^2}$, 
equation (\ref{Z1}) is rewritten as  
\begin{align}
(Z-1)
&= -  Z\lambda \int_0^\xi \frac{d\rho}{1-\rho}
\big(\rho G_{0\rho}+ G_{0\rho}'\big)\;,\qquad \text{or}
\label{ZZ1}
\\
Z^{-1} 
&= 1+\lambda \int_0^\xi d\rho 
\frac{G_{0\rho}}{1-\rho}
- \lambda \int_0^\xi d\rho 
\Big( G_{0\rho} - \frac{G_{0\rho}'}{1-\rho}\Big)\;.
\label{ZZ2}
\end{align}
We insert (\ref{ZZ1}) into the last term of the first line of
(\ref{G-alphabeta}) and divide by $Z$:
\begin{align}
G_{\alpha\beta} 
&= Z^{-1} + \frac{\lambda}{Z^{-1}} 
G_{\alpha\beta} \frac{(1-\alpha)(1-\beta)}{1-\alpha\beta}
\int_0^\xi \frac{\rho \,d\rho}{(1-\rho)^2} \Big( 
\frac{(1-\alpha)G_{\alpha\rho}}{1-\alpha \rho} 
- G_{0\rho} \Big)
  \nonumber
  \\
  &
+\lambda  G_{\alpha\beta} 
\frac{(1-\alpha)(1-\beta)}{1-\alpha\beta}
\int_0^\xi d\rho \Big(G_{0\rho} -\frac{G_{0\rho}'}{1-\rho}\Big)
\nonumber
  \\
  &
-\lambda (1-\alpha)(1-\beta)
\int_0^\xi \frac{d\rho}{(1-\rho)} 
\Big(\frac{\rho}{1-\beta \rho} \frac{G_{\beta\rho}}{\rho-\alpha}
- \frac{\alpha}{1-\beta \alpha} \frac{G_{\beta\alpha}}{\rho-\alpha}
\Big)\;.
\label{Gab-1}
\end{align}
Insertion of (\ref{ZZ2}) gives
\begin{align}
G_{\alpha\beta} &=  1 + \lambda \Bigg\{
\int_0^\xi d\rho \frac{G_{0\rho}}{1-\rho} 
+ \frac{\displaystyle G_{\alpha\beta} \frac{(1-\alpha)(1-\beta)}{1-\alpha\beta}
\int_0^\xi \frac{\rho \,d\rho}{(1-\rho)^2} \Big( 
\frac{(1-\alpha)G_{\alpha\rho}}{1-\alpha \rho} 
- G_{0\rho} \Big)}{\displaystyle
1 +\lambda \int_0^\xi d\rho 
\frac{G_{0\rho}}{1-\rho}
- \lambda \int_0^\xi d\rho 
\Big( G_{0\rho} - \frac{G_{0\rho}'}{1-\rho}\Big)}
  \nonumber
  \\
  &
\qquad +  \Big(
\frac{(1-\alpha)(1-\beta)}{1-\alpha\beta}G_{\alpha\beta}  -1\Big)
\int_0^\xi d\rho \Big(G_{0\rho} -\frac{G_{0\rho}'}{1-\rho}\Big)
\nonumber
  \\
  &
\qquad -(1-\alpha)(1-\beta)
\int_0^\xi \frac{d\rho}{(1-\rho)} 
\Big(\frac{\rho}{1-\beta \rho} \frac{G_{\beta\rho}}{\rho-\alpha}
- \frac{\alpha}{1-\beta \alpha} \frac{G_{\beta\alpha}}{\rho-\alpha}
\Big)\Bigg\}\;.
\label{Gab-2}
\end{align}
Rational fraction expansion yields
\begin{align}
G_{\alpha\beta} &=  1 + \lambda \Bigg\{
G_{\alpha\beta} \frac{(1-\beta)}{1-\alpha\beta} 
\Big( \frac{ 
(1-\alpha) \mathcal{K}^\xi_\alpha -\alpha \mathcal{X}^\xi 
+ \mathcal{M}^\xi_\alpha -\mathcal{L}^\xi_\alpha }{
1+ \lambda (\mathcal{X}^\xi-\mathcal{Y}^\xi) }
- \alpha \ln(1-\xi)\Big)
\nonumber
\\
& 
\qquad +  \Big(
\frac{(1-\alpha)(1-\beta)}{1-\alpha\beta}G_{\alpha\beta}  -1\Big)
\mathcal{Y}^\xi
\nonumber
  \\
  &
\qquad +\frac{ (1-\alpha)}{1-\alpha\beta}\big(
\mathcal{M}^\xi_\beta-\mathcal{L}^\xi_\beta\big)
-\frac{\alpha (1-\beta)}{1-\alpha\beta}\big( \mathcal{L}^\xi_\beta+
\mathcal{N}_{\alpha\beta}^\xi\big) 
\Bigg\}\;,
\label{G-KLMN}
\end{align}
where 
\begin{align}
\mathcal{K}_\alpha^\xi &:=\int_0^\xi d\rho \; 
\frac{G_{\alpha\rho} -G_{0\rho}}{(1-\rho)^2} \;,
&
\mathcal{L}_\alpha^\xi &:=\int_0^\xi d\rho \; 
\frac{G_{\alpha\rho} -G_{0\rho}}{(1-\rho)} \;,
\\
\mathcal{M}_\alpha^\xi &:=\int_0^\xi d\rho \; 
\frac{\alpha\,G_{\alpha\rho}}{(1-\alpha \rho)} \;,
&
\mathcal{N}_{\alpha\beta}^\xi &:=\int_0^\xi d\rho \; 
\frac{G_{\rho\beta}-G_{\alpha\beta}}{(\rho -\alpha)} \;,
\\
\mathcal{X}^\xi &:= 
\int_0^\xi d\rho \frac{G_{0\rho}}{(1-\rho)}\;,
&
\mathcal{Y}^\xi &:= 
\int_0^\xi d\rho \Big(G_{0\rho}- \frac{G'_{0\rho}}{(1-\rho)}\Big)\;.
\label{calKLMN}
\end{align}
The functions $\mathcal{K}^\xi_\alpha,\mathcal{X}^\xi,\ln(1-\xi)$ are singular
for $\xi \to 1$. Fortunately, these singularities cancel.  For that we
evaluate (\ref{G-KLMN}) separately for $\alpha=0$ and $\beta=0$:
\begin{align}
G_{0\beta} &= 1+\lambda \Big(((1-\beta)G_{0\beta}-1)\mathcal{Y}^\xi
+ \mathcal{M}^\xi_\beta-\mathcal{L}^\xi_\beta  \Big)\;,
\\
G_{\alpha 0} &= 1+\lambda \Big( G_{\alpha 0} 
\Big\{ \frac{ 
(1-\alpha) \mathcal{K}_\alpha^\xi -\alpha \mathcal{X}^\xi 
+ \mathcal{M}_\alpha^\xi -\mathcal{L}_\alpha^\xi }{
1+ \lambda (\mathcal{X}^\xi-\mathcal{Y}^\xi) }
- \alpha \ln(1-\xi)\Big\}
\nonumber
\\
& \qquad +((1-\alpha) G_{\alpha 0}-1)\mathcal{Y}^\xi
- \alpha \mathcal{N}_{\alpha 0}^\xi \Big)\;.
\label{Ga0-2}
\end{align}
Taking the symmetry $G_{\alpha 0}=G_{0\alpha}$ into account, 
the term in braces in (\ref{Ga0-2}) must be equal to 
$\mathcal{M}^\xi_\alpha-\mathcal{L}^\xi_\alpha+ 
\alpha \mathcal{N}^\xi_{\alpha 0}$, so that
(\ref{G-KLMN}) becomes 
\begin{align}
G_{\alpha\beta}
&=  1 
+ \lambda \Bigg( 
\frac{1-\beta}{1-\alpha\beta} 
\frac{G_{\alpha\beta}}{G_{0\alpha}} \big(
\mathcal{M}_\alpha^\xi-\mathcal{L}^\xi_\alpha 
+ \alpha \mathcal{N}^\xi_{\alpha 0}\big) 
+  \Big( \frac{(1-\alpha)(1-\beta)}{1-\alpha \beta} 
G_{\alpha\beta}-1\Big) \mathcal{Y}^\xi
\nonumber
\\
&
+\frac{1-\alpha}{1-\alpha\beta} 
\big(\mathcal{M}_\beta^\xi-\mathcal{L}_\beta^\xi\big)
-\frac{\alpha(1-\beta)}{1-\alpha\beta}  
\big(\mathcal{L}_{\beta}^\xi+\mathcal{N}^\xi_{\alpha\beta}\big)
\Bigg)\;.
\label{G-LMNY}
\end{align}
We have checked the equality between (\ref{G-KLMN}) and (\ref{G-LMNY}) 
perturbatively up to second order in $\lambda$; actually we discovered
it in this way. 

Since the model is renormalisable \cite{Grosse:2004yu}, 
the limit $\xi \to 1$ can be taken. We have thus proven:

\begin{Theorem}
  The renormalised planar connected two-point function
  $G_{\alpha\beta}$ of self-dual noncommutative $\phi^4_4$-theory
  (with continuous indices) satisfies the integral equation
\begin{align}
G_{\alpha\beta}
&=  1 
+ \lambda \bigg( 
\frac{1-\alpha}{1-\alpha\beta} 
\big(\mathcal{M}_\beta-\mathcal{L}_\beta -\beta \mathcal{Y} \big)
+ \frac{1-\beta}{1-\alpha\beta} 
\big(\mathcal{M}_\alpha-\mathcal{L}_\alpha -\alpha \mathcal{Y}\big) 
\nonumber
\\
& \qquad 
+\frac{1-\beta}{1-\alpha\beta} 
\Big(\frac{G_{\alpha\beta}}{G_{0\alpha}} -1\Big)\big(
\mathcal{M}_\alpha-\mathcal{L}_\alpha + \alpha \mathcal{N}_{\alpha
  0}\big) 
-\frac{\alpha(1-\beta)}{1-\alpha\beta}  
\big(\mathcal{L}_{\beta}+\mathcal{N}_{\alpha\beta}-
\mathcal{N}_{\alpha 0}\big)
\nonumber
\\
&\qquad +  \frac{(1-\alpha)(1-\beta)}{1-\alpha \beta} 
(G_{\alpha\beta}-1)\mathcal{Y}
\bigg)\;,
\label{G-LMNY-neu}
\end{align}
where $\alpha,\beta \in [0,1)$,
\begin{align}
\mathcal{L}_\alpha &:=\int_0^1 \!\! d\rho \; 
\frac{G_{\alpha\rho} -G_{0\rho}}{1-\rho} \;,
&
\mathcal{M}_\alpha &:=\int_0^1 \!\! d\rho \; 
\frac{\alpha\,G_{\alpha\rho}}{1-\alpha \rho} \;,
&
\mathcal{N}_{\alpha\beta} &:=\int_0^1 \!\! d\rho \; 
\frac{G_{\rho\beta}-G_{\alpha\beta}}{\rho -\alpha} \;,
\end{align}
and $\mathcal{Y}=\lim_{\alpha\to 0}
\frac{\mathcal{M}_\alpha-\mathcal{L}_\alpha}{\alpha}$. 
\end{Theorem}

\section{Perturbative solution}

The integral equation (\ref{G-LMNY-neu}) is the starting point of 
a perturbative solution $G_{\alpha\beta}=\sum_{n=0}^\infty \lambda^n 
G_{\alpha\beta}^{(n)}$. This gives directly the renormalised planar
two-point function, without need of Feynman graph computation and
further renormalisation steps. In particular, all integrals in 
$\mathcal{L}_{\alpha},\mathcal{M}_{\alpha\beta},
\mathcal{N}_{\alpha\beta}$ are regular (explicitly verified to
$\mathcal{O}(\lambda^4)$). The solution is conveniently expressed in terms of
\emph{iterated integrals} labelled by \emph{rooted trees}:
\begin{align}
I_\alpha  := \int_0^1 dx \;\frac{\alpha}{1-\alpha x} &=
-\ln(1-\alpha)\;,  
\nonumber
\\
I_{\genfrac{}{}{0pt}{}{\alpha}{\bullet}} := 
 \int_0^1 dx \;\frac{\alpha \,I_x}{1-\alpha x} &=
\mathrm{Li}_2(\alpha)+\frac{1}{2} \big(\ln(1-\alpha)\big)^2
\nonumber
\\
I_{\genfrac{}{}{0pt}{}{\alpha}{\bullet~\bullet}}
=  \int_0^1 dx \;\frac{\alpha \,I_x\cdot I_x}{1-\alpha x} &=
-2 \,\mathrm{Li}_3\Big(- \frac{\alpha}{1-\alpha}\Big)\;,
\nonumber
\\
I_{\genfrac{}{}{0pt}{}{\alpha}{\genfrac{}{}{0pt}{}{\bullet}{\bullet}}}
=  \int_0^1 dx \;\frac{\alpha \,
I_{\genfrac{}{}{0pt}{}{x}{\bullet}}}{1-\alpha x} &=
-2\mathrm{Li}_3\Big(- \frac{\alpha}{1-\alpha}\Big)
-2\mathrm{Li}_3(\alpha)
- \ln(1-\alpha)\zeta(2) 
\nonumber
\\
& +\ln(1-\alpha) \mathrm{Li}_2(\alpha)
+\frac{1}{6} \big(\ln(1-\alpha)\big)^3\;.
\label{iteratedintegral}
\end{align}
Similar iterated integrals appeared in toy models for the Hopf algebra of
Connes-Kreimer \cite{Connes:1998qv} (where the root is above). 
We find up to third order
\begin{align}
G_{\alpha\beta} &= 1 + \lambda \Big\{ 
A(I_\beta-\beta) + B(I_\alpha-\alpha)\Big\}
\nonumber
\\
& + \lambda^2\Big\{
A 
\big(\beta I_{\genfrac{}{}{0pt}{}{\beta}{\bullet}}-\beta I_\beta \big) 
- \alpha A B \big((I_\beta)^2-2\beta I_\beta+I_\beta \big)
\nonumber
\\
&\quad +B
\big(\alpha I_{\genfrac{}{}{0pt}{}{\alpha}{\bullet}}-\alpha I_\alpha\big) 
-\beta  B A \big((I_\alpha)^2-2\alpha I_\alpha + I_\alpha \big)
\nonumber
\\
& \quad + AB
\big((I_{\genfrac{}{}{0pt}{}{\alpha}{\bullet}}-\alpha) +
(I_{\genfrac{}{}{0pt}{}{\beta}{\bullet}} -\beta)
+(I_\alpha-\alpha)(I_\beta-\beta)
+\alpha\beta(\zeta(2)+1)\big) 
\Big\}  
\nonumber
\\
& + \lambda^3 \Big\{
A \mathcal{W}_{\beta} + 
\alpha AB  \big(-\mathcal{U}_{\beta} +I_\alpha I_\beta +
I_{\di{\alpha}{\bullet}} I_\beta\big)
+ \alpha  A^2B ( \mathcal{V}_{\beta})
\nonumber
\\
&\quad + B \mathcal{W}_{\alpha}
+\beta B A \big(-\mathcal{U}_{\alpha} 
+I_\beta I_\alpha +I_{\di{\beta}{\bullet}} I_\alpha\big)
+ \beta B^2 A (\mathcal{V}_{\alpha})
\nonumber
\\
& 
\quad +
A B
\big( \mathcal{T}_\beta +\mathcal{T}_\alpha 
- I_\beta (I_\alpha)^2 - I_\alpha (I_\beta)^2
-6 I_\alpha I_\beta \big) 
\nonumber
\\
& \quad + AB^2\big((1-\alpha)
(I_{\genfrac{}{}{0pt}{}{\alpha}{\bullet}}-\alpha) +3 I_\alpha I_\beta
+I_{\di{\beta}{\bullet}} I_\alpha+ I_\beta (I_\alpha)^2 \big)
\nonumber
\\
&\quad +  B A^2 \big((1-\beta)(I_{\genfrac{}{}{0pt}{}{\beta}{\bullet}} -\beta)
+3 I_\alpha I_\beta 
+I_{\di{\alpha}{\bullet}} I_\beta+ I_\alpha (I_\beta)^2  
\big)\Big\} + \mathcal{O}(\lambda^4)\;,
\label{Gab}
\end{align}
where we have defined
\begin{align}
A &:=\frac{1-\alpha}{1-\alpha\beta}\;,\qquad
B:=\frac{1-\beta}{1-\alpha\beta}\;,
\nonumber
\\
\mathcal{T}_\beta &:= 
\beta I_{\genfrac{}{}{0pt}{}{\beta}{\bullet}}-\beta I_\beta 
+(I_\beta-\beta)\;,
\nonumber
\\
\mathcal{U}_\beta 
&:=
- \beta  I_{\di{\beta}{\bullet~\bullet}} 
-(I_\beta)^3
+ \beta I_{\di{\beta}{\bullet}} I_\beta
+ 2 I_{\di{\beta}{\bullet}}I_\beta 
+ \beta \zeta(2) I_\beta
- 2\beta  \zeta(3) 
\nonumber
\\
&-2 (I_\beta)^2
+ \beta (I_\beta)^2
+I_{\di{\beta}{\bullet}}
+\beta I_{\genfrac{}{}{0pt}{}{\beta}{\bullet}}
+ 2 I_\beta -\beta^2 \;,
\nonumber
\\
\mathcal{V}_\beta 
&:=  \beta I_{\di{\beta}{\di{\bullet}{\bullet}}} 
- \beta^2  I_{\di{\beta}{\bullet~\bullet}} 
- 2\beta^2  \zeta(3) 
+ 2 \beta I_{\di{\beta}{\bullet}}I_\beta 
-  I_\beta^3
+ 2 \beta I_\beta \zeta(2) 
- 3 \beta^2  \zeta(2) 
\nonumber
\\
&+ (1-\beta)\big(2 \beta I_{\di{\beta}{\bullet}} 
- 3 I_\beta^2 + 3  \beta I_\beta -3  I_\beta +\beta\big)\,,
\nonumber
\\
\mathcal{W}_\beta &:= 
(I_{\di{\beta}{\di{\bullet}{\bullet}}}
- \beta \zeta(2))
-\frac{1}{2} I_\beta \frac{I_\beta-\beta}{\beta}
+\frac{1}{2} (I_\beta)^2 
-(I_{\di{\beta}{\bullet}}-\beta)
-\frac{1}{2}(I_\beta-\beta)
-\frac{1}{2}\beta^2\;.
\end{align}
We notice that up to third order, the solution $G_{\alpha\beta}$ is a
polynomial with rational coefficients in $\alpha$, $\beta$, $A$, $B$,
$\zeta(2)$, $\zeta(3)$ and the iterated integrals\footnote{There
  appears the integral $\displaystyle \frac{I_\alpha-\alpha}{\alpha} =
  \int_0^1 d\rho\; \frac{\alpha\rho}{1-\alpha\rho}$, which seems to be
  more appropriate than $I_\alpha$ itself.} (\ref{iteratedintegral}).
It is remarkable how the non-symmetric equation (\ref{G-LMNY-neu}) leads 
to the symmetric solution for $G_{\alpha\beta}$!

It is tempting to conjecture that $G_{\alpha\beta}$ is at \emph{any
  order $n$} a polynomial with rational coefficients in
$\alpha,\beta$, $A$, $B$, (multiple) zeta values
\cite{Kontsevich:2001} and iterated integrals labelled by rooted trees
with at most $n$ vertices. Proving this conjecture is a main step to
prove Borel summability of the two-point function. Note that there are
$n!$ (not necessarily connected) rooted trees (with multiplicities)
with $n$ vertices, which means that at order $n$ in the perturbation
series there would be only $\mathcal {O}(n!)$ independent
contributions.

We show in the next section for $n=4$ that the corresponding
Schwinger-Dyson equation for an $(n>2)$-point function is
\emph{linear} and inhomogeneous, with the inhomogeneity given by
$m$-point functions with $m<n$. Such equations are
straightforward to estimate if the two-point function is known. After
all, this would be the very first construction of an interacting
quantum field theory in four dimensions.

\section{Four-point Schwinger-Dyson equation}

Here we demonstrate for the planar four-point function that the
knowledge of the two-point function permits a successive construction
of the whole theory. Starting point is the Schwinger-Dyson equation
for the planar connected four-point function $G_{abcd}$.  Following
the $a$-face in direction of the arrow, there is a distinguished
vertex at which the first $ab$-line starts. For this vertex there are
two possibilities for the matrix index of the diagonally opposite corner to the 
$a$-face: either $c$ or a summation vertex $p$:
\begin{align}
\parbox{28mm}{\begin{picture}(28,20)
\put(0,10){\xy <1cm,0cm>:
(0,0) *++={ }; (1,0) *++={ } **\frm{o},
(0.55,0)="Bi",
(1.3,0)="Bo",
(-0.55,0)="Ai",
(-1.3,0)="Ao",
(0,0.55)="Ci",
(0,1.3)="Co",
(0,-0.55)="Di",
(0,-1.3)="Do",
\ar @{=} "Ai";"Ao"
\ar @{{ }<} "Ai";"Ao" <3pt> 
\ar @{{ }>} "Ai";"Ao" <-3pt>
\ar @{=} "Bi";"Bo"
\ar @{{ }<} "Bi";"Bo" <3pt> 
\ar @{{ }>} "Bi";"Bo" <-3pt>
\ar @{=} "Ci";"Co"
\ar @{{ }<} "Ci";"Co" <3pt> 
\ar @{{ }>} "Ci";"Co" <-3pt>
\ar @{=} "Di";"Do"
\ar @{{ }<} "Di";"Do" <3pt> 
\ar @{{ }>} "Di";"Do" <-3pt>
\endxy}
\put(9,0){\mbox{\scriptsize$b$}}
\put(16,0){\mbox{\scriptsize$c$}}
\put(16,20){\mbox{\scriptsize$d$}}
\put(9,20){\mbox{\scriptsize$a$}}
\put(2,13){\mbox{\scriptsize$a$}}
\put(2,6){\mbox{\scriptsize$b$}}
\put(22,13){\mbox{\scriptsize$d$}}
\put(22,6){\mbox{\scriptsize$c$}}
\end{picture}}
&= 
\parbox{42mm}{\begin{picture}(40,36)
\put(2,17){\xy <1cm,0cm>:
(0.4,0.8) *++={ }; (0.9,0.8) *++={ } **\frm{o},
(-0.7,-1) *++={ }; (-0.2,-1) *++={ } **\frm{o},
(-1.55,0) *++={ }; (-1.05,0) *++={ } **\frm{o},
(0.75,0.8)="Bi",
(1.4,0.8)="Bo",
(-0.8,0)="AAi",
(-1.2,0)="AAo",
(-1.9,0)="Ai",
(-2.5,0)="Ao",
(0.4,1.2)="Ci",
(0.4,1.8)="Co",
(-0.7,-0.1)="Di",
(-0.7,-0.6)="Do",
(-0.7,-1.3)="Ei",
(-0.7,-1.9)="Eo",
(-0.7,0.1)="Fi",
(-0.7,0.8)="Fm",
(0.05,0.8)="Fo",
(-0.6,0)="Gi",
(0.4,0)="Gm",
(0.4,0.45)="Go",
\ar @{=} "Ai";"Ao"
\ar @{{ }<} "Ai";"Ao" <3pt> 
\ar @{{ }>} "Ai";"Ao" <-3pt>
\ar @{=} "Bi";"Bo"
\ar @{{ }<} "Bi";"Bo" <3pt> 
\ar @{{ }>} "Bi";"Bo" <-3pt>
\ar @{=} "Ci";"Co"
\ar @{{ }<} "Ci";"Co" <3pt> 
\ar @{{ }>} "Ci";"Co" <-3pt>
\ar @{=} "Di";"Do"
\ar @{=} "Ei";"Eo"
\ar @{{ }<} "Ei";"Eo" <3pt> 
\ar @{{ }>} "Ei";"Eo" <-3pt>
\ar @`{"Fm"} @{=} "Fi";"Fo"
\ar @`{"Gm"} @{=} "Gi";"Go"
\ar @{=} "AAi";"AAo"
\endxy}
\put(16,-2){\mbox{\scriptsize$b$}}
\put(22,-2){\mbox{\scriptsize$c$}}
\put(33,33){\mbox{\scriptsize$d$}}
\put(28,33){\mbox{\scriptsize$a$}}
\put(4,19){\mbox{\scriptsize$a$}}
\put(4,13){\mbox{\scriptsize$b$}}
\put(16,13){\mbox{\scriptsize$b$}}
\put(22,13){\mbox{\scriptsize$c$}}
\put(16,19){\mbox{\scriptsize$a$}}
\put(37,27){\mbox{\scriptsize$d$}}
\put(37,21){\mbox{\scriptsize$c$}}
\end{picture}}
+ 
\parbox{55mm}{\begin{picture}(50,30)
\put(3,15){\xy <1cm,0cm>:
(0.75,0) *++={ }; (1.5,0) *++={ } **\frm{o},
(2.2,0) *++={ }; (2.7,0) *++={ } **\frm{o},
(1.5,1.1) *++={ }; (2,1.1) *++={ } **\frm{o},
(1.5,-1.1) *++={ }; (2,-1.1) *++={ } **\frm{o},
(-1.55,0) *++={ }; (-1.05,0) *++={ } **\frm{o},
(-0.8,0)="AAi",
(-1.2,0)="AAo",
(1.15,0)="Hi",
(1.9,0)="Ho",
(2.5,0)="Bi",
(3,0)="Bo",
(-1.9,0)="Ai",
(-2.5,0)="Ao",
(0.97,0.35)="CCi",
(1.3,0.8)="CCo",
(-0.7,-0.1)="Di",
(-0.7,-0.6)="Do",
(1,-0.32)="EEi",
(1.3,-0.8)="EEo",
(-0.7,0.1)="Fi",
(-0.7,0.7)="Fm",
(0.2,0.7)="Fn",
(0.5,0.35)="Fo",
(-0.7,-0.1)="Di",
(-0.7,-0.9)="Dm",
(0.2,-0.8)="Dn",
(0.52,-0.33)="Do",
(-0.6,0)="Gi",
(0.35,0)="Gm",
(1.85,1.15)="Ci",
(2.5,1.15)="Co",
(1.85,-1.15)="Ei",
(2.5,-1.15)="Eo",
\ar @{=} "Ai";"Ao"
\ar @{{ }<} "Ai";"Ao" <3pt> 
\ar @{{ }>} "Ai";"Ao" <-3pt>
\ar @{=} "Bi";"Bo"
\ar @{{ }<} "Bi";"Bo" <3pt> 
\ar @{{ }>} "Bi";"Bo" <-3pt>
\ar @{=} "Ci";"Co"
\ar @{{ }<} "Ci";"Co" <3pt> 
\ar @{{ }>} "Ci";"Co" <-3pt>
\ar @{=} "Ei";"Eo"
\ar @{{ }<} "Ei";"Eo" <3pt> 
\ar @{{ }>} "Ei";"Eo" <-3pt>
\ar @`{"Dm","Dn"} @{=} "Di";"Do"
\ar @`{"Fm","Fn"} @{=} "Fi";"Fo"
\ar @{=} "Gi";"Gm"
\ar @{=} "Hi";"Ho"
\ar @{=} "CCi";"CCo"
\ar @{=} "EEi";"EEo"
\ar @{=} "AAi";"AAo"
\endxy}
\put(49,0){\mbox{\scriptsize$b$}}
\put(49,5){\mbox{\scriptsize$c$}}
\put(49,22.5){\mbox{\scriptsize$d$}}
\put(49,28.5){\mbox{\scriptsize$a$}}
\put(5,17){\mbox{\scriptsize$a$}}
\put(5,11){\mbox{\scriptsize$b$}}
\put(17,17){\mbox{\scriptsize$a$}}
\put(17,11){\mbox{\scriptsize$b$}}
\put(56,17){\mbox{\scriptsize$d$}}
\put(56,11){\mbox{\scriptsize$c$}}
\put(24,11){\mbox{\scriptsize$p$}}
\end{picture}}
\label{SD-main}
\end{align}
We let $G^{(1)}_{abcd}$ and $G^{(2)}_{abcd}$ be the corresponding two 
graphs on the rhs.  We write $G^{(1)}_{abcd}$ as a product of the vertex
$Z^2\lambda$, the left connected two-point function, the downward
two-point function and an insertion, which is reexpressed by means of
the Ward-identity:
\begin{align}
G^{(1)}_{abcd}  &= Z^2 \lambda G_{ab} 
G_{bc} G^{ins}_{[ac]d} 
= Z \lambda G_{ab}  G_{bc} \frac{1}{(|a|-|c|)} \big( G_{cd} - 
G_{ad}\big) 
\nonumber 
\\
&=Z \lambda G_{ab}  G_{bc} G_{cd} G_{da}
\frac{1}{(|a|-|c|)} \Big( \frac{1}{G_{ad}} - 
\frac{1}{G_{cd}}\Big)\;.
\end{align}

In the last graph in (\ref{SD-main}) we open the $p$-face to get an
insertion. However, this insertion is not into the full connected
four-point function! The connected four-point function $G_{abcd}$ contains 
at least one $ab$-line, which is not present in the subgraph under 
consideration. Therefore, we have to subtract from the general 
four-point insertion the insertion 
into the $G_{ab}$ two-point function:
\begin{align}
G^{(2)}_{abcd} &=  Z^2 \lambda \parbox{30mm}{\begin{picture}(30,20)
\put(3,10){\xy <1cm,0cm>:
(0.75,0) *++={ }; (1.5,0) *++={ } **\frm{o},
(1.15,0)="Bi",
(1.9,0)="Bo",
(-0.4,0)="Gi",
(0.35,0)="Gm",
\ar @{=} "Bi";"Bo"
\ar @{{ }<} "Bi";"Bo" <3pt> 
\ar @{{ }>} "Bi";"Bo" <-3pt>
\ar @{=} "Gi";"Gm"
\ar @{<{ }} "Gi";"Gm" <3pt> 
\ar @{>{ }} "Gi";"Gm" <-3pt>
\endxy}
\put(5,12){\mbox{\scriptsize$a$}}
\put(5,6){\mbox{\scriptsize$b$}}
\put(25,12){\mbox{\scriptsize$a$}}
\put(25,6){\mbox{\scriptsize$b$}}
\end{picture}}  
\times \sum_p 
 \parbox{55mm}{\begin{picture}(50,30)
\put(3,15){\xy <1cm,0cm>:
(0.75,0) *++={ }; (1.5,0) *++={ } **\frm{o},
(2.2,0) *++={ }; (2.7,0) *++={ } **\frm{o},
(1.5,1.1) *++={ }; (2,1.1) *++={ } **\frm{o},
(1.5,-1.1) *++={ }; (2,-1.1) *++={ } **\frm{o},
(0,-1.08) *++={ }; (-0.6,-1.08) *++={ } **\frm{o},
(1.15,0)="Hi",
(1.9,0)="Ho",
(2.5,0)="Bi",
(3,0)="Bo",
(-0.35,-1.1)="Ai",
(-1,-1.1)="Ao",
(0.97,0.35)="CCi",
(1.3,0.8)="CCo",
(-0.7,-0.1)="Di",
(-0.7,-0.6)="Do",
(1,-0.32)="EEi",
(1.3,-0.8)="EEo",
(-0.7,0.1)="Fi",
(-0.7,0.7)="Fm",
(0.2,0.7)="Fn",
(0.5,0.35)="Fo",
(-0.7,-0.1)="Di",
(-0.7,-0.9)="Dm",
(0.2,-0.8)="Dn",
(0.52,-0.33)="Do",
(-0.6,0)="Gi",
(0.35,0)="Gm",
(1.85,1.15)="Ci",
(2.5,1.15)="Co",
(1.85,-1.15)="Ei",
(2.5,-1.15)="Eo",
\ar @{=} "Ai";"Ao"
\ar @{{ }<} "Ai";"Ao" <3pt> 
\ar @{{ }>} "Ai";"Ao" <-3pt>
\ar @{=} "Bi";"Bo"
\ar @{{ }<} "Bi";"Bo" <3pt> 
\ar @{{ }>} "Bi";"Bo" <-3pt>
\ar @{=} "Ci";"Co"
\ar @{{ }<} "Ci";"Co" <3pt> 
\ar @{{ }>} "Ci";"Co" <-3pt>
\ar @{=} "Ei";"Eo"
\ar @{{ }<} "Ei";"Eo" <3pt> 
\ar @{{ }>} "Ei";"Eo" <-3pt>
\ar @{=} "Dn";"Do"
\ar @`{"Fm","Fn"} @{=} "Fi";"Fo"
\ar @{=} "Gi";"Gm"
\ar @{=} "Hi";"Ho"
\ar @{=} "CCi";"CCo"
\ar @{=} "EEi";"EEo"
\endxy}
\put(2,0){\mbox{\scriptsize$b$}}
\put(4,6){\mbox{\scriptsize$p$}}
\put(34,0){\mbox{\scriptsize$b$}}
\put(36,5){\mbox{\scriptsize$c$}}
\put(34,22.5){\mbox{\scriptsize$d$}}
\put(36,28.5){\mbox{\scriptsize$a$}}
\put(2.5,17){\mbox{\scriptsize$a$}}
\put(7,11.5){\mbox{\scriptsize$p$}}
\put(42,17){\mbox{\scriptsize$d$}}
\put(41,11){\mbox{\scriptsize$c$}}
\end{picture}}
\nonumber
\\[2ex]
&=  Z^2 \lambda \parbox{30mm}{\begin{picture}(30,20)
\put(3,10){\xy <1cm,0cm>:
(0.75,0) *++={ }; (1.5,0) *++={ } **\frm{o},
(1.15,0)="Bi",
(1.9,0)="Bo",
(-0.4,0)="Gi",
(0.35,0)="Gm",
\ar @{=} "Bi";"Bo"
\ar @{{ }<} "Bi";"Bo" <3pt> 
\ar @{{ }>} "Bi";"Bo" <-3pt>
\ar @{=} "Gi";"Gm"
\ar @{<{ }} "Gi";"Gm" <3pt> 
\ar @{>{ }} "Gi";"Gm" <-3pt>
\endxy}
\put(5,12){\mbox{\scriptsize$a$}}
\put(5,6){\mbox{\scriptsize$b$}}
\put(25,12){\mbox{\scriptsize$a$}}
\put(25,6){\mbox{\scriptsize$b$}}
\end{picture}}  
\sum_p \left( \parbox{35mm}{\begin{picture}(35,20)
\put(2,9){{\xy <1cm,0cm>:
(0,0) *++={ }; (1,0) *++={ } **\frm{o},
(0.46,0.31)="Bi",
(1.1,0.81)="Bo",
(0.46,-0.31)="Ci",
(1.1,-0.81)="Co",
(0.1,0.55)="Fi",
(0.2,1.3)="Fo",
(-0.05,-0.55)="Gi",
(-0.15,-1.3)="Go",
(-.39,0.41)="Ai",
(-1.1,1.11)="Ao",
(-0.39,-0.41)="Di",
(-1.1,-1.2)="Do",
(-1.5,0.04)="E1",
(-1.5,-0.04)="E2",
(-1.58,0.12)="E1a",
(-1.58,-0.12)="E2a",
(-1.65,0)="E0",
\ar @{=} "Bi";"Bo"
\ar @{{ }<} "Bi";"Bo" <3pt>
\ar @{{ }>} "Bi";"Bo" <-3pt>
\ar @{=} "Ci";"Co"
\ar @{{ }<} "Ci";"Co" <3pt>
\ar @{{ }>} "Ci";"Co" <-3pt>
\ar @{=} "Fi";"Fo"
\ar @{{ }<} "Fi";"Fo" <3pt>
\ar @{{ }>} "Fi";"Fo" <-3pt>
\ar @{=} "Gi";"Go"
\ar @{{ }<} "Gi";"Go" <3pt>
\ar @{{ }>} "Gi";"Go" <-3pt>
\ar @`{"Do"} @{=} "Di";"E2"
\ar @`{"Ao"} @{=} "Ai";"E1"
\ar @{<{ }} "E1";"Ao" <3pt>
\ar @{<{ }} "E2";"Do" <3pt>
\ar @`{"E0"} @{.} "E1a";"E2a"
\endxy}}
\put(1,12){\mbox{\scriptsize$a$}}
\put(1,4){\mbox{\scriptsize$p$}}
\put(22,20){\mbox{\scriptsize$d$}}
\put(16,21){\mbox{\scriptsize$a$}}
\put(26,18){\mbox{\scriptsize$d$}}
\put(28,13){\mbox{\scriptsize$c$}}
\put(25,-2){\mbox{\scriptsize$b$}}
\put(28,4){\mbox{\scriptsize$c$}}
\put(19,-4){\mbox{\scriptsize$b$}}
\put(12,-2){\mbox{\scriptsize$p$}}
\end{picture}} 
- \hspace*{-2em}
\parbox{35mm}{\begin{picture}(35,36)
\put(2,20){{\xy <1cm,0cm>:
(0,0) *++={ }; (1,0) *++={ } **\frm{o},
(-0.2,-1.5) *++={ }; (0.6,-1.5) *++={ } **\frm{o},
(0.46,0.31)="Bi",
(1.1,0.81)="Bo",
(0.46,-0.31)="Ci",
(1.1,-0.81)="Co",
(0.1,0.55)="Fi",
(0.2,1.3)="Fo",
(-0.05,-0.55)="Gi",
(-0.12,-1)="Go",
(-.5,-1.09)="Ai",
(-1.1,-.39)="Ao",
(-.51,-1.89)="Di",
(-1.1,-2.7)="Do",
(-1.5,-1.46)="E1",
(-1.5,-1.54)="E2",
(-1.58,-1.38)="E1a",
(-1.58,-1.62)="E2a",
(-1.65,-1.5)="E0",
(0.3,-1.5)="Hi",
(1,-1.5)="Ho",
\ar @{=} "Bi";"Bo"
\ar @{{ }<} "Bi";"Bo" <3pt>
\ar @{{ }>} "Bi";"Bo" <-3pt>
\ar @{=} "Ci";"Co"
\ar @{{ }<} "Ci";"Co" <3pt>
\ar @{{ }>} "Ci";"Co" <-3pt>
\ar @{=} "Fi";"Fo"
\ar @{{ }<} "Fi";"Fo" <3pt>
\ar @{{ }>} "Fi";"Fo" <-3pt>
\ar @{=} "Gi";"Go"
\ar @{=} "Hi";"Ho"
\ar @{{ }<} "Hi";"Ho" <3pt>
\ar @{{ }>} "Hi";"Ho" <-3pt>
\ar @`{"Do"} @{=} "Di";"E2"
\ar @`{"Ao"} @{=} "Ai";"E1"
\ar @{<{ }} "E1";"Ao" <3pt>
\ar @{<{ }} "E2";"Do" <3pt>
\ar @`{"E0"} @{.} "E1a";"E2a"
\endxy}}
\put(1,8){\mbox{\scriptsize$a$}}
\put(1,0){\mbox{\scriptsize$p$}}
\put(25,1){\mbox{\scriptsize$p$}}
\put(25,7){\mbox{\scriptsize$b$}}
\put(24,11.5){\mbox{\scriptsize$b$}}
\put(29,14){\mbox{\scriptsize$c$}}
\put(28,23){\mbox{\scriptsize$c$}}
\put(27,30){\mbox{\scriptsize$d$}}
\put(22,30){\mbox{\scriptsize$d$}}
\put(16,31){\mbox{\scriptsize$a$}}
\put(19,11){\mbox{\scriptsize$b$}}
\put(13.5,12){\mbox{\scriptsize$a$}}
\end{picture}} 
\right).
\end{align}
In the very last graph, the whole $ab$-line is considered as part of 
the lower bubble, giving the insertion $G^{ins}_{[ap]b}$. The
remaining upper bubble has the two-point function $G_{ab}$ amputated, 
but together with the $G_{ab}$ prefactor in front of the sum we obtain the
full connected four-point function. In summary, we have 
\begin{align}
G^{(2)}_{abcd}  
&= Z^2 \lambda\Big( \sum_p G_{ab} G^{ins}_{[ap]bcd} - 
 G^{ins}_{[ap]b} G_{abcd}\Big)
\nonumber
\\
&= Z \lambda \sum_p G_{ab} \frac{1}{|a|-|p|} 
\big(G_{pbcd} - G_{abcd} \big)
 \nonumber
\\
&-Z \lambda  \sum_p\frac{1}{|a|-|p|} 
\big( G_{pb} - G_{ab}\big) G_{abcd}
\nonumber
\\
&= Z \lambda \sum_p \frac{1}{|a|-|p|} 
\big(G_{ab} G_{pbcd} - G_{pb}  G_{abcd}\big)\;.
\end{align}
After amputation of the external two-point functions we obtain the
Schwinger-Dyson equation for the \emph{renormalised} 1PI four-point
function $G_{abcd}= G_{ab} G_{bc} G_{cd} G_{da} \Gamma_{abcd}^{ren}$
as follows:
\begin{align}
\Gamma_{abcd}^{ren}=  Z \lambda 
\frac{1}{|a|-|c|} \Big( \frac{1}{G_{ad}} - 
\frac{1}{G_{cd}}\Big)+ Z \lambda \sum_p \frac{1}{|a|-|p|} 
G_{pb} \Big( \frac{G_{dp}}{G_{ad}} \Gamma^{ren}_{pbcd} 
-  \Gamma^{ren}_{abcd}\Big)\;.
\label{SD-4}
\end{align}
In terms to the 1PI function (\ref{G-inverse}) we have
\begin{align}
Z^{-1}  \Gamma_{abcd}^{ren} &=  \lambda 
\Big( 1- \frac{\Gamma_{ad}^{ren}- \Gamma_{cd}^{ren}}{|a|-|c|}\Big)
\nonumber
\\[-2ex]
&+ \lambda \sum_p 
\frac{|a|+|d|+\mu^2-\Gamma^{ren}_{ad}}{|p|+|b|+\mu^2
-\Gamma^{ren}_{pb}}
\frac{ \dfrac{\Gamma^{ren}_{pbcd} -\Gamma^{ren}_{abcd}}{|p|-|a|}}{
|p|+|d|+\mu^2-\Gamma^{ren}_{pd}}
\nonumber
\\
& + \lambda \Gamma_{abcd}^{ren} 
\sum_p 
\frac{1-\dfrac{\Gamma^{ren}_{ad}-\Gamma^{ren}_{pd}}{|a|-|p|} }{
(|p|+|b|+\mu^2-\Gamma^{ren}_{pb})(
|p|+|d|+\mu^2-\Gamma^{ren}_{pd})}\;.
\end{align}
Passing to the integral representation and the variables (\ref{alpha}),
we find for $\Gamma_{\alpha\beta\gamma\delta} := \Gamma_{abcd}^{ren}$  
\begin{align}
Z^{-1} \Gamma_{\alpha\beta\gamma\delta}
&=  \lambda 
\Big( 1- \frac{(1-\gamma)\Gamma_{\alpha\delta}- (1-\alpha)
\Gamma_{\gamma\delta}}{(1-\delta)(\alpha-\gamma)}\Big)
\nonumber
\\
&+ \lambda \int_0^\xi \rho \,d\rho
\frac{(1-\beta)(1-\alpha\delta-\Gamma_{\alpha\delta})}{
(1-\beta\rho-\Gamma_{\beta\rho})}
\frac{ \dfrac{\Gamma_{\rho\beta\gamma\delta} 
-\Gamma_{\alpha\beta\gamma\delta}}{\rho-\alpha}}{
1-\delta\rho- \Gamma_{\delta\rho}}
\nonumber
\\
& + \lambda \Gamma_{\alpha\beta\gamma\delta}
 \int_0^\xi \frac{\rho \,d\rho}{(1-\rho)}
\frac{(1-\beta)
\Big((1-\delta)-\dfrac{(1-\rho)\Gamma_{\alpha\delta}- (1-\alpha)
\Gamma_{\rho\delta}}{(\alpha-\rho)}\Big)
}{
(1-\beta \rho -\Gamma_{\beta\rho})(1-\delta\rho -\Gamma_{\delta\rho})}
\nonumber
\\
&=\lambda 
\Big(  \frac{1}{G_{\alpha\delta}}
- \frac{(1-\alpha)(1-\gamma\delta) (G_{\alpha\delta}-G_{\gamma\delta})
}{
G_{\alpha\delta}G_{\gamma\delta} (1-\delta)(\alpha-\gamma)}\Big)
\nonumber
\\
&+ \lambda \int_0^\xi \rho \,d\rho
\frac{(1-\beta)(1-\alpha\delta) G_{\beta\rho} G_{\delta\rho} }{
G_{\alpha\delta}(1-\beta\rho)(1-\delta\rho)}
\frac{\Gamma_{\rho\beta\gamma\delta} 
-\Gamma_{\alpha\beta\gamma\delta}}{\rho-\alpha}
\nonumber
\\
& - \lambda \Gamma_{\alpha\beta\gamma\delta}
 \int_0^\xi \rho \,d\rho
\frac{(1-\beta)(1-\alpha\delta)G_{\beta\rho}}{
G_{\alpha\delta}(1-\beta \rho)(1-\delta\rho )}
\frac{ (G_{\rho\delta} -G_{\alpha\delta})}{(\rho-\alpha)}
\nonumber
\\
& 
+ \lambda \Gamma_{\alpha\beta\gamma\delta}
 \int_0^\xi d\rho 
\Big( \frac{G_{\beta\rho}}{1-\rho}
- \frac{\beta G_{\beta\rho}}{1-\beta\rho }
- \frac{ G_{\beta\rho}(1-\beta)}{(1-\delta\rho)(1-\beta\rho )}
\Big)\;.
\end{align}
Now we insert (\ref{ZZ2}) for $Z^{-1}$ and bring the last two lines to
the lhs. It arises a combination where the limit $\xi \to 1$ exists: 

\clearpage

\begin{Theorem}
  The renormalised planar 1PI four-point function
  $\Gamma_{\alpha\beta\gamma\delta}$ of self-dual noncommutative
  $\phi^4_4$-theory (with continuous indices $\alpha,\beta,\gamma,\delta 
\in [0,1)$)  satisfies the integral equation
\begin{align}
\Gamma_{\alpha\beta\gamma\delta}
&
=\lambda \cdot 
\frac{\begin{array}{l}
\displaystyle \Big( 1 
- \frac{(1-\alpha)(1-\gamma\delta) (G_{\alpha\delta}-G_{\gamma\delta})
}{G_{\gamma\delta} (1-\delta)(\alpha-\gamma)}
\\ \displaystyle
\qquad \qquad +  \int_0^1 \rho \,d\rho
\frac{(1-\beta)(1-\alpha\delta) G_{\beta\rho} G_{\delta\rho} }{
(1-\beta\rho)(1-\delta\rho)}
\frac{\Gamma_{\rho\beta\gamma\delta} 
-\Gamma_{\alpha\beta\gamma\delta}}{\rho-\alpha}\Big)
\end{array}}{
\begin{array}{l}
\displaystyle 
G_{\alpha\delta}
+ \lambda \Big((\mathcal{M}_\beta-\mathcal{L}_\beta-\mathcal{Y})
G_{\alpha\delta}
+  \int_0^1 d\rho 
\frac{ G_{\alpha\delta} G_{\beta\rho}(1-\beta)}{(1-\delta\rho)(1-\beta\rho )}
\\
\qquad\qquad
\displaystyle + 
 \int_0^1 \rho \,d\rho
\frac{(1-\beta)(1-\alpha\delta)G_{\beta\rho}}{
(1-\beta \rho)(1-\delta\rho )}
\frac{ (G_{\rho\delta} -G_{\alpha\delta})}{(\rho-\alpha)}\Big)
\end{array}
}
\end{align}
\end{Theorem}

In lowest order we find
\begin{align}
\Gamma_{\alpha\beta\gamma\delta}
=\lambda 
-\lambda^2 \Big(
&\frac{(1-\gamma) (I_\alpha-\alpha) 
- (1-\alpha)(I_\gamma-\gamma)}{\alpha-\gamma}
\nonumber
\\
&+ \frac{(1-\delta)(I_\beta-\beta)-(1-\beta)(I_\delta-\delta)}{\beta-\delta} 
\Big) + \mathcal{O}(\lambda^3)\;.
\end{align}
Note that $\Gamma_{\alpha\beta\gamma\delta}$ is cyclic in the four
indices, and that $\Gamma_{0000}=\lambda +  \mathcal{O}(\lambda^3)$.

\section*{Acknowledgements}

R.W. thanks the Erwin-Schr\"odinger-Institute in Vienna for invitation
and hospitality in connection with the Senior Research Fellowship in
spring 2009.  H.G. thanks the Mathematical Institute of the WWU
M\"unster for a number of invitations and hospitality and support from
the SFB 478. Most of the work was done during these mutual visits.
We also thank the EU-NCG network MRTN-CT-2006-031962.

\end{document}